\documentclass[twocolumn,english,american,prb]{revtex4}
\usepackage[T1]{fontenc}
\usepackage[latin1]{inputenc}
\usepackage{amsmath}
\usepackage{graphicx}
\usepackage{amssymb}

\makeatletter

\providecommand{\tabularnewline}{\\}

\usepackage{babel}
\makeatother
\begin{document}
\selectlanguage{english}
\newcommand{\hexar} {   \begin{picture}(20,15)(-4,5)     \includegraphics[width=0.5cm]{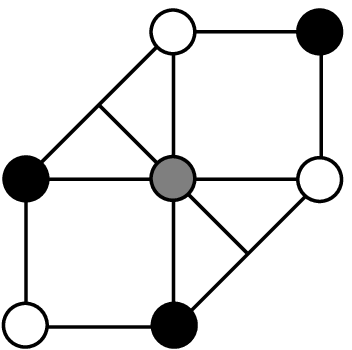}   \end{picture} }

\newcommand{\hexbr} {   \begin{picture}(20,15)(-4,5)     \includegraphics[width=0.5cm]{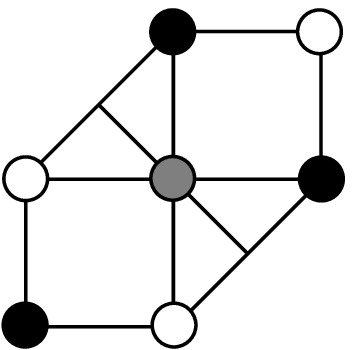}   \end{picture} }

\newcommand{\hexal} {   \begin{picture}(20,15)(-4,5)     \includegraphics[width=0.5cm]{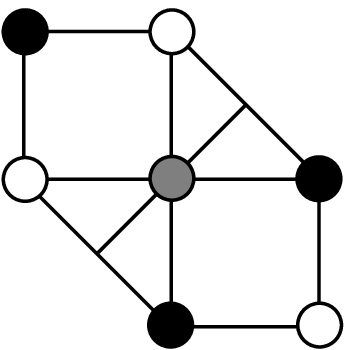}   \end{picture} }

\newcommand{\hexbl} {   \begin{picture}(20,15)(-4,5)     \includegraphics[width=0.5cm]{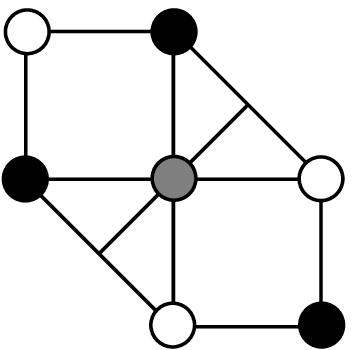}   \end{picture} }

\newcommand{\hexo} {   \begin{picture}(10,5)(0,2)     \includegraphics[width=0.4cm]{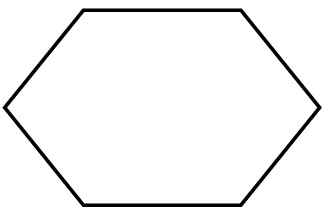} \end{picture} }

\selectlanguage{american}
\begin{abstract}
We study the dynamical properties of spinless fermions on the checkerboard
lattice. Our main interest is the limit of large nearest-neighbor
repulsion $V$ as compared with hopping $|t|$. The spectral functions
show broad low-energy excitation which are due to the dynamics of
fractionally charged excitations. Furthermore, it is shown that the
fractional charges contribute to the electrical current density. 
\end{abstract}

\title{Spectral functions and optical conductivity of spinless fermions
on a checkerboard lattice}

\author{F. Pollmann, P. Fulde}

\address{Max-Planck-Institut für Physik komplexer Systeme, Nöthnitzer Strasse
38, 01187 Dresden, Germany}

\author{E. Runge}

\address{Technische Universität Ilmemau, Fakultät für Mathematik und Naturwissenschaften,
FG Theoretische Physik I, 98693 Ilmenau, Germany }

\maketitle

\section{Introduction}

Excitations with fractional charge have fascinated physicists for
some time. An early investigation by Su, Schrieffer and Heeger \cite{su1979}
dates back to 1979 and deals with the chain molecule trans-polyacetylene
$\mathrm{(CH)_{n}}$. This polymer has two ground states A and B with
alternating single and double bonds between $\mathrm{C}$ atoms. In
one of them the double bond is on the right side of a given $\mathrm{C}$
atom while in the other ground state it is on the left side. A soliton
or kink excitation has the property of an interface between the two
ground states, i.e., on one side of the soliton we have alternating
bonds like in ground state A and on the other side the bonds are arranged
like in ground state B. Shifting the soliton along the chain does
not change the energy. Su et al. showed that depending on the occupational
number of the kink state which can be either empty, singly or doubly
occupied, the excitation has a spin or charge only. In other words,
there exist excitations with spin-charge separation but no excitations
with fractional charge yet. This changes when trans-polyacetylene
is heavily doped. It turns out that at certain doping concentrations
of either particles or holes there are not only excitations with charge
and spin separated but their charge is a fraction $\nu$ of the electronic
charge $e$ only.\cite{su1980} In the simplest case $\nu=\pm1/3,\ \pm2/3$
etc. It should be mentioned that this phenomenon does not require
interactions between electrons but occurs within a single- (or independent)
particle description. It involves lattice degrees of freedom though,
since a double $\mathrm{C=C}$ bond has a different length than a
single $\mathrm{C-C}$ bond. Doped trans-polyacetylene serves as an
example of a one-dimensional system with fractional charges. 

Fractional charges are also found in two dimensions in the much celebrated
fractional quantum Hall effect (FQHE). It was Laughlin\cite{laughlin1983}
who introduced this concept here. He explained with it the behavior
of electrons in high-quality $\mathrm{GaAs/AlGaAs}$ heterostructures
in an applied magnetic field $B$ normal to the planes. When the field
deviates slightly from $B=nhc/e\nu$, where $n$ is the density of
the electrons and $\nu$ is a rational fraction with odd denominator,
excitations with charge $\pm\nu e$ occur. As usual $h$ is Planck's
constant. The excitations with fractional charge $\nu e$ are based
on a new type of correlated ground state called the Laughlin state.
It has been also demonstrated that the excitations fulfill fractional
statistics, i.e., when two of them are exchanged the phase changes
by $e^{i\nu\pi}$. In distinction to the previously considered trans-polyacetylene,
electron-electron interactions are crucial here. In an applied magnetic
field, correlations become strong since the kinetic energy of the
electrons is reduced.

The question was left open whether or not fractionally charged excitations
exist in two or three dimensional systems without magnetic fields.
In Ref.~[\onlinecite{fulde2002}] it was suggested that in a pyrochlore
lattice, a prototype of a three dimensional structure with geometrical
frustration, excitations with charge $e/2$ do exist. The pyrochlore
structure is found, for example, in the transition metal compound
$\mathrm{LiV_{2}O_{4}}$. In that compound, vanadium has a half-integer
valency and electronic correlations are strong as indicated by a large
$\gamma$ coefficient in the low temperature specific heat $C=\gamma T$.
Although $\mathrm{LiV_{2}O_{4}}$ has motivated work of the kind presented
here, we do not claim that the theory of fractional charges applies
to that specific material. Generally, prerequisites for fractional
charges in the considered model are strong short-range correlations
and certain band fillings. In order to study a system of this sort
in more detail, a checkerboard lattice is considered which also allows
for numerical investigations. The checkerboard lattice can be thought
of as a projection of the pyrochlore lattice onto a plane. Although
there are differences in the physics of the two systems due to the
different dimensions, one can learn much from the simpler two dimensional
system. Numerous studies of the pyrochlore and checkerboard lattice
were done and reported in Refs.~[\onlinecite{misguich2003,hermele2004,lauchli2004,diep2005}]
for spin degrees of freedom and [\onlinecite{shannon2004, zhang2005, zhou2005}]
for charge degrees of freedom. The question has not been posed yet
how one would be able to detect these excitations with fractional
charges in case they exist. It is the purpose of the present investigation
to shed some light on that topic by calculating spectral functions
and optical conductivity of fully spin polarized or, equivalently,
spinless fermions on a checkerboard lattice. It is demonstrated that
the fractionalization of charge leads to characteristic features which
are absent when fractionalization is forbidden. Perhaps the present
investigation may sharpen the attention of experimentalists when doing
photoemission experiments on, e.g., systems with spinel structure
when the filling factors are right. Also the experimental progress
in the generation of optical lattices should be noted. Recently it
has been reported that 3D optical lattices can by generated and filled
with either bosons or fermions to simulate relevant Hamiltonians,
e.g., in Ref.~[\onlinecite{hofstetter2002}] and citations therein.

The paper is organized as follows. First we introduce in Section \ref{sec:Hamiltonian-and-lattices}
the model Hamiltonian that is used for our calculations and show how
under certain conditions fractionally charged particles (fcp's) arise
on a checkerboard lattice. In Section \ref{sec:Numerical-details},
we briefly dicuss the numerical details and introduce the approximations
we applied. The calculated spectral functions and optical conductivities
are presented in Section \ref{sec:Results}. Finally, the results
and conclusions are summarized in Section \ref{sec:Summary-and-outlook}.

\section{Hamiltonian and lattices\label{sec:Hamiltonian-and-lattices}}

\begin{figure}
\begin{center}\begin{tabular}{ccc}
(a)\includegraphics[%
  width=17mm]{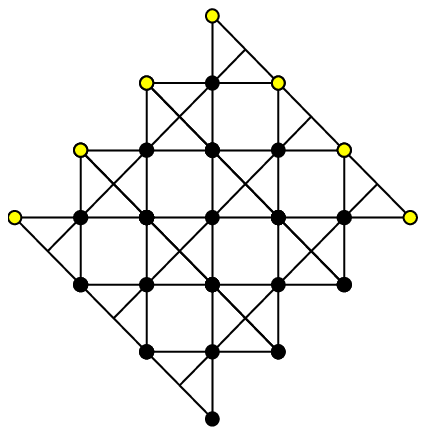}&
(b)\includegraphics[%
  width=23mm]{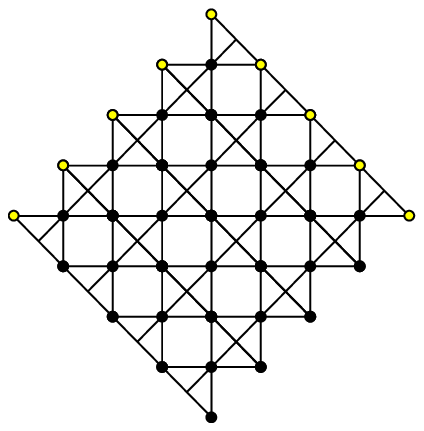}&
(c)\includegraphics[%
  width=30mm]{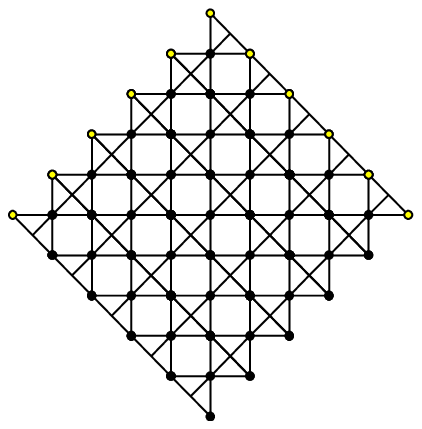}\tabularnewline
\end{tabular}\end{center}

\caption{Checkerboard lattice (periodic boundary conditions are used for the
calculations). Lattices with (a) 18 sites, (b) 32 sites and (c) 50
sites are considered.\label{cap:Checkerboard-lattice}}
\end{figure}

The model Hamiltonian 

\begin{equation}
H=-t\sum_{\langle i,j\rangle}\left(c_{i}^{\dag}c_{j}^{\vphantom{\dag}}+\mbox{h.c.}\right)+V\sum_{\langle i,j\rangle}n_{i}n_{j}\label{eq:hamil}\end{equation}
describes spinless fermions on a checkerboard lattice (see Fig.~\ref{cap:Checkerboard-lattice})
 with nearest-neighbor hopping amplitude $t$ and nearest-neighbor
repulsion $V$. The operators $c_{i}^{\vphantom{\dag}}(c_{i}^{\dag})$
annihilate (create) a particle on site $i$. The density operator
is $n_{i}=c_{i}^{\dag}c_{i}^{\vphantom{\dag}}$. To avoid boundary
effects, which are expected to be strong, we choose periodic boundary
conditions in actual calculations. 

The band structure of non-interacting ($V=0$) particles on the crisscrossed
checkerboard lattice consists of a dispersive and a flat band \begin{eqnarray}
\varepsilon^{-}(\mathbf{k}) & = & -2t-4t\cos\frac{k_{x}a}{\sqrt{2}}\cos\frac{k_{y}a}{\sqrt{2}}\nonumber \\
\varepsilon^{+}(\mathbf{k}) & = & 2t,\label{eq:disp_checker}\end{eqnarray}
where $a$ is the lattice constant. In passing, we mention that in
the presence of interactions correlations are generally strong in
the flat band regimes. The kinetic energy does not counteract optimization
of repulsions in that regime.\cite{mielke1991} In the following we
assume $t>0$. Note that our coordinate system is rotated by $45°$
relative to that of, e.g., Ref.~{[}\onlinecite{runge2004}{]}. The
resulting difference in boundary conditions can lead to noticeable
numerical differences in particular for very small cluster sizes. 

Our main interest is in the regime $t/V\ll1$, at half filling. When
$t=0$ the ground state manifold is macroscopically degenerate: Every
configuration with exactly two particles on each crisscrossed square
is a ground state. This is the two dimensional equivalent of the tetrahedron
rule \cite{anderson1956} (or Anderson rule). We refer to configurations
fulfilling the tetrahedron rule as {}``allowed configurations''.
The macroscopic degeneracy is lifted for small but finite hopping
amplitude $t$. 

\begin{figure}
\begin{center}\begin{tabular}{cc}
\includegraphics[%
  width=30mm]{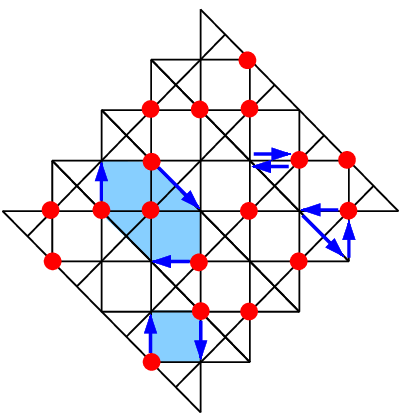}\hspace{-30mm}(a)\hspace{30mm}~&
\includegraphics[%
  width=30mm]{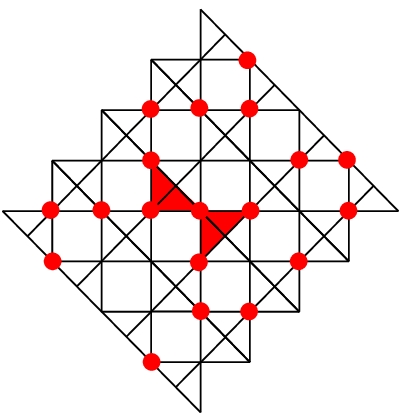}\hspace{-30mm}(b)\hspace{30mm}~\tabularnewline
\includegraphics[%
  width=30mm]{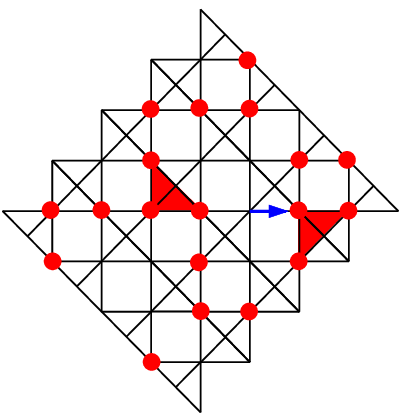}\hspace{-30mm}(c)\hspace{30mm}~&
\includegraphics[%
  width=30mm]{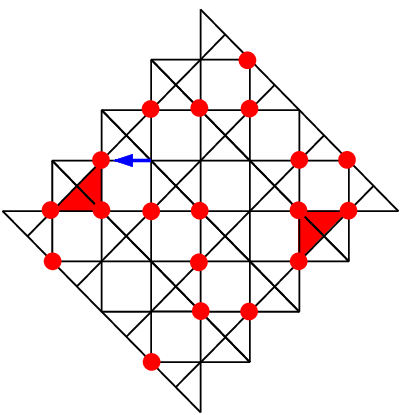}\hspace{-30mm}(d)\hspace{30mm}~\tabularnewline
\end{tabular}\end{center}

\caption{(color online) (a) Example of an allowed configuration of a checkerboard
lattice at half filling with possible low-order hopping processes.
(b)-(d) Adding one particle leads to two mobile defects, marked by
triangles. Arrows mark motions of particles.\label{cap:Checkerboard-lattice-doped}}
\end{figure}

In the following we discuss processes up to order $t^{3}/V^{2}$.
They can be classified as self-energy contributions and ring exchanges
(see Fig.~\ref{cap:Checkerboard-lattice-doped}(a)).%
 The self-energy contributions are identical for all allowed configurations
and therefore the ground state degeneracy is not lifted by them. The
total amplitude of ring exchange around empty squares is proportional
to $t^{2}/V$ and vanishes for fermions. However, the macroscopic
degeneracy is lifted by ring exchanges $\sim t^{3}/V^{2}$ around
hexagons. Note that for fermions the signs of the corresponding matrix
elements depend on the sequence in which the sites are enumerated.\cite{runge2004} 

Placing one additional particle with charge $e$ onto an empty site
leads to a violation of the tetrahedron rule on two adjacent crisscrossed
squares (see Fig.~\ref{cap:Checkerboard-lattice-doped}(b)). The
energy is increased by $4V$ since the added particle has four nearest-neighbors.
Particles on a crisscrossed square with three particles can hop to
another crisscrossed square without creating more violations of the
tetrahedron rule, i.e., without increase of repulsive energy (see
panels (c) and (d) in Fig.~\ref{cap:Checkerboard-lattice-doped}).
By these processes, two local defects (violating the tetrahedron rule)
can separate over large distances thereby gaining additional kinetic
energy of order $t$. This is observed in our calculations. We will
see below in the context of Fig.~\ref{cap:Integrated-spectral-density_3x3}
that the bandwidth from an artificially restricted calculation in
which the defects are not allowed to separate is much smaller than
that from the unrestricted calculation. 

These findings motivate the introduction of the following effective
Hamiltonian that acts only on the subspace of allowed configurations
with a given number of violations (e.g., two violations in Fig.~\ref{cap:Checkerboard-lattice-doped}
(b)-(d)):

\begin{eqnarray}
H_{\mbox{\tiny eff}} & = & g\sum_{\small\{\hexo\}}(-1)^{n_{0}}\left(|\hexar\rangle\langle\hexbr|+|\hexal\rangle\langle\hexbl|+\mbox{h.c.}\right)\nonumber \\
 & - & t\sum_{\langle i\  j\rangle}\mathcal{P}\left(c_{i}^{\dag}c_{j}^{\vphantom{\dag}}+\mbox{h.c.}\right)\mathcal{P}\ .\label{eq:Heff}\end{eqnarray}
It includes the lowest order ring exchange process with $g\sim t^{3}/V^{2}$
(compare Ref.~\onlinecite{runge2004}). We will refer to it as the
t-g model. Note that the sign of the matrix elements depends on the
configuration, i.e, for row wise enumeration of sites a process has
a minus sign whenever the site in the center of the corresponding
hexagon is occupied ( $n_{0}=1$) and a plus sign when it is empty.
The hopping $t$ is projected onto the manifold of allowed configurations
plus fixed numbers of violations by the projector $\mathcal{P}$.
By introducing the independent parameters $t$ and $g$ the effect
of ring exchange onto the dynamics of fractional excitations can be
estimated. The tendency for confinement can be studied even on rather
small clusters by increasing the ring exchange strength $g$. 

One can expect that for an infinitely large system the two defects
can propagate freely and independently throughout the system. In that
case they should be considered as two elementary excitations each
having a dispersion $\tilde{\varepsilon}(\mathbf{k})\sim t$ and each
carrying the fractional charge $e/2$. Similarly, the repulsion-energy
increase of $4V$ is split between the two elementary excitations.
We will refer to these excitations as fractionally charged particles
(fcp's). It should be emphasized that they are not quasiparticles
in the sense of Landau's Fermi liquid theory because they are not
adiabatically connected to independent fermions. Analogously, taking
one particle out of the system generates two fractionally charged
holes (fch's), i.e., two crisscrossed square with one particle only.
Furthermore, quantum fluctuations generate fcp and fch pairs. Starting
from an allowed configuration, when one particle is hopping to an
empty neighboring site it creates one crisscrossed square with three
particles (fcp) and one crisscrossed square with one particle (fch)
only.

Whether or not two defects actually \emph{do} separate completely
is a very subtle problem. The answer will depend on lattice type,
dimensionality, spin vs. spinless particles, etc. The answer may also
differ for zero and finite temperatures $T$ and depend on defect
concentration. In any case, the key observation is that whenever two
defects with charge $\pm e/2$ separate they change the background
configuration. Infinite defect separation (deconfinement) is more
plausible for liquidlike ground states than, for ground states with,
e.g., broken translational invariance. In the latter case, restoring
forces occur which can lead to linear confinement, analogously to
quark confinement known from QCD. Thus a thorough analysis of the
ground state of the undoped, i.e., half filled system is a vital step
towards answering the above questions. First numerical results for
an effective model on finite systems up to $64$ sites have been presented
in Ref.~{[}\onlinecite{runge2004}{]}. A two-fold degenerate ground
state with stripe order was found for the particular model on the
checkerboard lattice. Nevertheless, various observations suggested
that it would disappear in the thermodynamic limit. Recent calculations
and analytical considerations\cite{pollmann2006} indicate that this
supposition, i.e, that the order would disappear in the thermodynamic
limit, was erroneous. The presence of an ordered ground state leads
probably to weak confinement of fcp's with a characteristic length
of $L$ lattice spacings. A simple estimate is given by $Lg\sim t$,
i.e, $L\sim t/g=V^{2}/t^{2}$ lattice spacings. For the used parameters,
this leads to a large characteristic length of several hundred lattice
spacings (weak confinement). Note that the 3D pyrochlore lattice appears
at present to have a deconfined phase in which the fcp's can separate\cite{fulde2002,moessner2003a,hermele2004}.
The present work focuses on the dynamics of a weakly doped system
(half filling plus one extra particle) by calculating the spectral
functions and optical conductivity. We stress that the numerical simulations
are limited to small clusters. As a consequence, as long as clusters
of size $L\sim t/g=V^{2}/t^{2}\gg1$ are out of reach for the exact
diagonalization, we can not hope to find simple finite size scaling
relations towards the true thermodynamic limit. The present study
should be considered as a contribution towards the understanding of
the local behavior of fcp's. In particular it is impossible to distinguish
between a free or weakly bound pair of fcp's.

\section{Numerical details\label{sec:Numerical-details}}

Obviously, conventional approximation schemes such as mean-field theories
and Green's function decoupling schemes are not able to describe the
strong local correlation expressed by the tetrahedron rule. Furthermore,
until now neither a creation operator formalism nor a field theoretical
description for the fcp's have been derived. This suggests numerical
studies. Unfortunately, the fact that we are dealing with fermions
rules out the use of standard Monte-Carlo techniques. Thus we have
chosen the exact diagonalization of the Hamiltonian (\ref{eq:hamil})
for small finite lattices, even though the numerical effort increases
exponentially with system size. 

Diagonalization within the full Hilbert space was done for a lattice
containing 18 sites. Larger systems can be considered if we restrict
ourselves within the corresponding Hilbert spaces to certain low-energy
sectors. They are defined by the number of violations of the tetrahedron
rule. For the actual calculations we used the smallest possible Hilbert
spaces allowing for the dynamical processes that we are interested
in. These consists in the undoped case, i.e, at half filling of the
allowed configurations and those with one additional vacuum fluctuation
present (one fcp and one fch). In the doped case, i.e, with one particle
added to the system, the configurations account for no other violations
of the tetrahedron rule than those that are due to the added particle
(two fcp's). We refer to the space spanned by these selected configurations
as the {}``minimal Hilbert space'' in each case. The dimensions
are strongly reduced as compared with the full Hilbert space (see
Table \ref{cap:The-dimensions-of}).%
\begin{table}[h]
\begin{center}\begin{tabular}{lrrr}
&
$18$ sites&
$32$ sites&
$50$ sites\tabularnewline
\hline
\multicolumn{4}{l}{\emph{Half filling}}\tabularnewline
\hline
Full Hilbert space&
$48\ 620$&
$\quad601\ 080\ 390$&
$\quad1.2641\ 10^{14}$\tabularnewline
Allowed states &
$68$&
$2970$&
$67\ 832$\tabularnewline
Minimal (1 fluct.)&
$2\ 228$&
$168\ 858$&
$16\ 178\ 232$\tabularnewline
\hline
&
&
&
\tabularnewline
\hline
\multicolumn{4}{l}{\emph{System doped with one particle}}\tabularnewline
\hline
Full Hilbert space&
$43\ 758$&
$\quad565\ 722\ 720$&
$\quad1.2155\ 10^{14}$\tabularnewline
Minimal (0 fluct.) &
$1\ 323$&
$98\ 784$&
$8\ 698\ 450$\tabularnewline
Confined (0 fluct.)&
$612$&
$47\ 520$&
$1\ 695\ 800$\tabularnewline
One extra fluctuation&
$11\ 475$&
$2\ 435\ 808$&
-~~~~~~\tabularnewline
\hline
\end{tabular}\end{center}

\caption{Dimensions of the full Hilbert space and some relevant subspaces
for different lattice sizes.\label{cap:The-dimensions-of} }
\end{table}
 Calculations for the doped system have also been done for two other
subspaces. In one case, the two fcp's subspace is extended to three
fcp and one fch in order to check the validity of the results for
the two fcp subspace. We refer to this Hilbert space as {}``one extra
fluctuation''. In the second case, we confine the two fcp's to adjacent
criss-cross squares, i.e., particles do not split into fractional
charges. Such calculations have the purpose to demonstrate that charge
fractionalization leads to qualitative different behavior.

The spectral functions and optical conductivity of interacting many-particle
systems are expectation values of the form\begin{eqnarray}
G(z)=\langle\psi_{0}|A\ \frac{1}{z-H}\  A^{\dag}|\psi_{0}\rangle\end{eqnarray}
 and can therefore conveniently be calculated numerically by the Lanczos
continued fraction method \cite{gagliano1987} or kernel polynomial
expansion.\cite{silver1996} We found essentially identical results
for both algorithms. However, the implementation of the Lanczos method
turned out to be slightly faster. We rewrite first $G(z)$ as

\begin{eqnarray}
G(z)=\langle\psi_{0}|A\  A^{\dag}|\psi_{0}\rangle\langle\phi_{0}|\ \frac{1}{z-H}\ |\phi_{0}\rangle,\end{eqnarray}
where\begin{eqnarray}
|\phi_{0}\rangle=\frac{A^{\dag}|\psi_{0}\rangle}{\sqrt{\langle\psi_{0}|AA^{\dag}|\psi_{0}\rangle}}.\end{eqnarray}
Then the state $|\phi_{0}\rangle$ is taken as starting vector to
generate iteratively with the Lanczos algorithm an orthogonal basis
for the Hamiltonian $H$. Using the tridiagonal form of the Hamiltonian
with respect to the Lanczos basis and Kramers' rule, the expectation
value can be easily rewritten in terms of diagonal elements $a_{n}$
and off-diagonal elements $b_{n}$ of the Hamiltonian in form of a
continued fraction\begin{eqnarray}
G(z)=\frac{\langle\psi_{0}|AA^{\dag}|\psi_{0}\rangle}{z-a_{0}-b_{1}^{2}\frac{1}{z-a_{1}-b_{2}^{2}\frac{1}{z-a_{2}\dots}}}.\end{eqnarray}
Well converged results were obtained already after several hundred
iterations. 

There is, of course, always the question remaining in how far results
for finite clusters are indications for the behavior of systems in
the thermodynamic limit. Since we cannot go beyond 50 sites, the reader
should be cautioned against over interpreting the results.

\section{Results \label{sec:Results}}

\subsection{Spectral functions}

Direct insight into the dynamics of a many-body system is provided
by the spectral function \begin{eqnarray}
A(\mathbf{k},\omega)=A^{-}(\mathbf{k},\omega)+A^{+}(\mathbf{k},\omega),\end{eqnarray}
which is the probability for adding (+) or removing (-) a particle
with momentum $\mathbf{k}$ and energy $\omega$ ($\hbar=1$) to the
system. This function can be directly related to angular-resolved
photoemission spectroscopy (ARPES). The particle contribution is defined
by\begin{eqnarray}
A^{+}(\mathbf{k},\omega) & = & \lim_{\eta\rightarrow0^{+}}-\frac{1}{\pi}\\
 &  & \times\mbox{Im}\langle\psi_{0}^{N}|c_{\mathbf{k}}\ \frac{1}{\omega+i\eta+E_{0}-H}\  c_{\mathbf{k}}^{\dag}|\psi_{0}^{N}\rangle\nonumber \end{eqnarray}
and the hole contribution by\begin{eqnarray}
A^{-}(\mathbf{k},\omega) & = & \lim_{\eta\rightarrow0^{+}}-\frac{1}{\pi}\\
 & \times & \mbox{Im}\langle\psi_{0}^{N}|\  c_{\mathbf{k}}^{\dag}\ \frac{1}{\omega+i\eta-E_{0}+H}\  c_{\mathbf{k}}|\psi_{0}^{N}\rangle.\nonumber \end{eqnarray}
Here $|\psi_{0}^{N}\rangle$ is the ground state of the system with
$N$ particles. A small value of $\eta=0.1t$ is used for Lorentzian
broadening. The operators $c_{\mathbf{k}}^{\vphantom{\dag}}$ ($c_{\mathbf{k}}^{\dag}$)
are obtained from the corresponding operators in real-space representation
\begin{eqnarray}
c_{\mathbf{k}}^{\dag}=\frac{1}{\sqrt{N_{\mathbf{k}}}}\sum_{j}e^{i\mathbf{r}_{j}\mathbf{k}}c_{j}^{\dag},\end{eqnarray}
where $N_{\mathbf{k}}$ denotes the number of $\mathbf{k}$~points
in the extended BZ and the sum is taken over all lattice sites. The
resulting integrated spectral density is \begin{eqnarray}
D(\omega)=\frac{1}{N_{\mathbf{k}}}\sum_{\mathbf{k}}A(\mathbf{k},\omega).\end{eqnarray}
For a system with a translationally invariant ground state or if an
average over all degenerate ground states is considered, $D(\omega)$
is conveniently calculated in real space representation as a local
expectation value 

\begin{figure}
\begin{center}\includegraphics[%
  width=80mm]{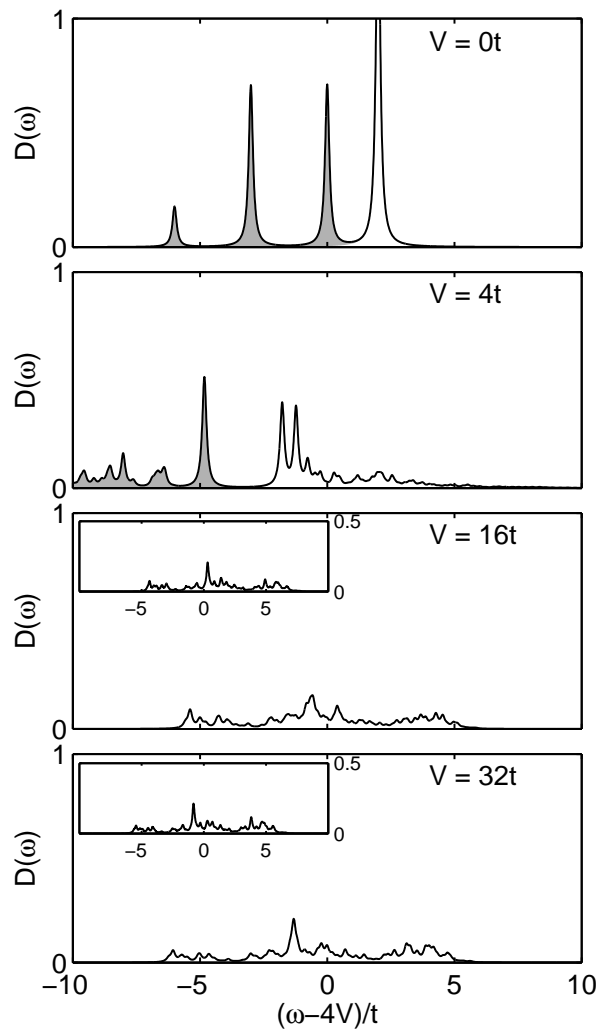}\end{center}

\caption{Integrated spectral density $D(\omega)$ of a $\sqrt{18}\times\sqrt{18}$
checkerboard lattice for increasing nearest-neighbor repulsion $V$.
The contributions of $A^{-}(\mathbf{k},\omega)$ is shown as shaded
area. The insets show $D(\omega)$ calculated for minimal Hilbert
spaces. Bandwidths and positions of the dominant features are almost
unchanged. The peaks are broadened by choosing $\eta=0.1t.$ \label{cap:Integrated-spectral-density_3x3}}
\end{figure}

The integrated spectral density $D(\omega)$ for the Hamiltonian (\ref{eq:hamil})
is displayed in Fig. \ref{cap:Integrated-spectral-density_3x3}.%
 The different values of the nearest neighbor repulsion $V$, show
the transition from an independent particle system to the strongly
correlated limit. A rather small system is considered in order to
allow for a diagonalization in the full Hilbert space with reasonable
computational effort. For $V=0$ the dispersion is given by Eq. \ref{eq:disp_checker}.
$D(\omega)$ includes contributions from the allowed $\mathbf{k}$-points
($\mathbf{k}=2\pi/3\ (n_{x},n_{y})\mbox{ with }(n_{x},n_{y})\in\mathbb{Z}^{2}$).
The dispersive band $\varepsilon^{-}$ is completely filled and contributes
only to $D^{-}(\omega)$ while the flat band is empty and contributes
exclusively to $D^{+}(\omega)$. 

Adding a particle increases the energy by $4V$. Removing it decreases
the repulsion energy by $2V$. Together this leads to an increasing
separation of particle and hole part of the spectrum with increasing
$V$ and finally to the formation of a gap. By this argument one expects
a metal-insulator transition as a function of $V$. At small but finite
$V$ the peaks are broadened as an incoherent background develops.
But we are interested in the opposite limit of large $V$. 

\begin{figure}
\begin{center}\includegraphics[%
  width=80mm]{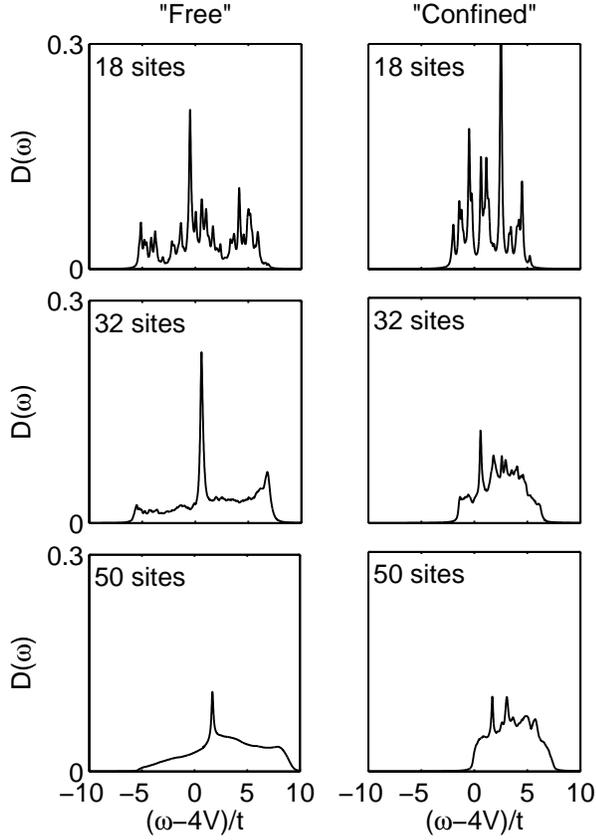}\end{center}

\caption{Integrated spectral density $D(\omega)$ for $V=25t$ calculated
for different lattice size. The left panel shows the data from {}``free''
systems which include states with separated fractional charges. The
right panel shows the data from systems where the two defect are {}``confined''
and can not separate. The peaks are broadened by choosing $\eta=0.1t.$
\label{cap:Integrated-spectral-density_V25t} }
\end{figure}

\begin{figure}
\begin{center}\includegraphics[%
  width=80mm]{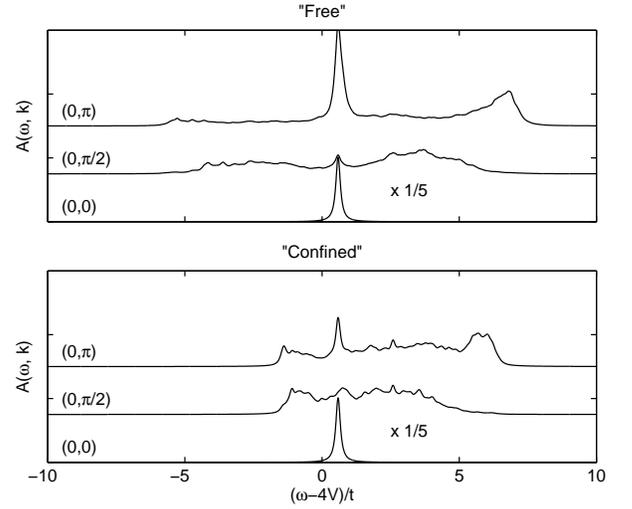}\end{center}

\caption{Spectral density $A(\mathbf{k},\omega)$ for $V=25t$ calculated
for some $k$-points. The upper panel shows data from the systems
which include states with separated fractional charges. The lower
panel shows the data from a system where the two defect are {}``confined''
and can not separate. The peaks are broadened by choosing $\eta=0.1t.$
\label{cap:Spectral-density_V25t} }
\end{figure}

In the limit of large $V$ it is sufficient to calculate the spectral
functions within the minimal Hilbert space, as is seen from the similarities
with results for the full Hilbert space, shown for parameter values
$V=16t$ and $V=32t$ in Fig. \ref{cap:Integrated-spectral-density_3x3}.
In particular, the bandwidth as well as the gross structure remain
unchanged. Therefore we may study much larger systems and compare
results for different lattice sizes. In particular, this enables us
to approach the question whether the two defects created by injecting
one particle are closely bound to each other or not. Unfortunately,
calculation on finite clusters can not distinguish truly between free
particles and weakly bound pairs. For a first answer, we compare for
different lattice sizes the results within the minimal Hilbert space
with those from an artificially restricted calculation keeping the
two defects confined to two adjacent crisscrossed squares. From Fig.~\ref{cap:Integrated-spectral-density_V25t}%
 it is seen that without the restriction a broad low-energy continuum
is obtained which is missing when the restriction is imposed. The
bandwidths are about $13t$ and $8t$ in the two cases. This suggests
a simple interpretation. The dynamics of two separate fcp's having
a bandwidth of $\approx6t$ each due to six nearest neighbors would
yield the calculated $13t$ while a confined added particle has a
much smaller bandwidth, i.e., $8t$. They are rather independent of
the system size (e.g., $\approx13t$ for the 18, 32 and 50 sites cluster
in the free case) and therefore expected to extrapolate to the thermodynamic
limit. A similar observation holds true for $A^{+}(\mathbf{k},\omega)$
(see Fig.~\ref{cap:Spectral-density_V25t}) where $\mathbf{k}=(0,\pi/2)$and
$\mathbf{k}=(0,\pi)$ are considered.%
 Additionally one observes two interesting facts: (i) The states with
large momentum $\mathbf{k}$ give the main contributions to the low-energy
continuum in $D(\omega)$. This is consistent with the picture that
a particle with large momentum more efficiently lowers its kinetic
energy when decaying into fractional excitations than a particle with
low momentum. (ii) $A^{+}(\mathbf{k},\omega)$ shows in the {}``confined''
calculation a peak-like feature at the low-energy edge that reminds
one of the Landau-Fermi liquid peak while nothing alike is seen for
the {}``free'' case. 

A third observation deserves a comment. For vanishing momentum, the
full spectral weight of $A^{+}$ is contained in a single sharp $\delta$-like
peak near the center of the band. This suggests that $|\tilde{\psi}^{N+1}\rangle=c_{\mathbf{k}=\mathbf{0}}^{\dag}|\psi_{0}\rangle$
is an exact eigenstate of the Hamiltonian $H$ in the limit $t/V\rightarrow0$
with $\tilde{H}|\tilde{\psi}^{N+1}\rangle=\tilde{E}|\tilde{\psi}^{N+1}\rangle$
as can be seen by evaluation of the following expression:\begin{eqnarray}
H|\tilde{\psi}^{N+1}\rangle & = & Hc_{\mathbf{k}=\mathbf{0}}^{\dag}|\psi_{0}\rangle\nonumber \\
 & = & \left[H,c_{\mathbf{k}=\mathbf{0}}^{\dag}\right]|\psi_{0}\rangle+c_{\mathbf{k}=\mathbf{0}}^{\dag}H|\psi_{0}\rangle\nonumber \\
 & = & \left[H,c_{\mathbf{k}=\mathbf{0}}^{\dag}\right]|\psi_{0}\rangle+E_{0}|\tilde{\psi}^{N+1}\rangle.\end{eqnarray}
 The contributions of the kinetic and the repulsive energy to the
commutator are given by\begin{eqnarray}
\left[H_{\mbox{\tiny kin}},c_{\mathbf{k}=\mathbf{0}}^{\dag}\right] & = & \left[\sum_{\mathbf{k^{\prime}}}\varepsilon(\mathbf{k^{\prime}})c_{\mathbf{k^{\prime}}}^{\dag}c_{\mathbf{k}^{\prime}}^{\vphantom{\dag}},c_{\mathbf{k}=\mathbf{0}}^{\dag}\right]\nonumber \\
 & = & \varepsilon(\mathbf{k}=\mathbf{0})c_{\mathbf{k}=\mathbf{0}}^{\dag}\end{eqnarray}
 and \begin{eqnarray}
\left[H_{\mbox{\tiny rep}},c_{\mathbf{k}=\mathbf{0}}^{\dag}\right] & = & \frac{V}{\sqrt{N}}\sum_{\langle ij\rangle}\left(c_{j}^{\dag}n_{i}^{\vphantom{\dag}}+c_{i}^{\dag}n_{j}^{\vphantom{\dag}}\right),\end{eqnarray}
respectively. The latter is calculated by using the real space representation
$c_{\mathbf{k}=\mathbf{0}}=1/\sqrt{N}\sum_{l}c_{l}^{\dag}$. Since
the ground state contains in the considered limit only configurations
that obey the tetrahedron rule, each empty site has exactly four occupied
neighbors. Thus the sum over all nearest neighbors applied to the
ground state leads to\begin{eqnarray}
\left[H_{\mbox{\tiny rep.}},c_{\mathbf{k}=\mathbf{0}}^{\dag}\right]|\psi_{0}\rangle & = & 4V\frac{1}{\sqrt{N}}\sum_{i}c_{i}^{\dag}|\psi_{0}\rangle\nonumber \\
 & = & 4V|\tilde{\psi}^{N+1}\rangle.\end{eqnarray}
Collecting everything yields\begin{eqnarray}
H|\tilde{\psi}^{N+1}\rangle=\left(\varepsilon(\mathbf{k}=\mathbf{0})+4V+E_{0}\right)|\tilde{\psi}^{N+1}\rangle.\end{eqnarray}
 The ground state wave function with one added particle of zero momentum
is an eigenfunction of the Hamiltonian with energy $\varepsilon(\mathbf{k}=\mathbf{0})+4V+E_{0}$.
Note that for the doped system no vacuum fluctuation is taken into
account and thus one has to add a constant energy shift from the self-energy
contributions (loss of ground state correlations).

\begin{figure}
\begin{center}\includegraphics[%
  width=80mm]{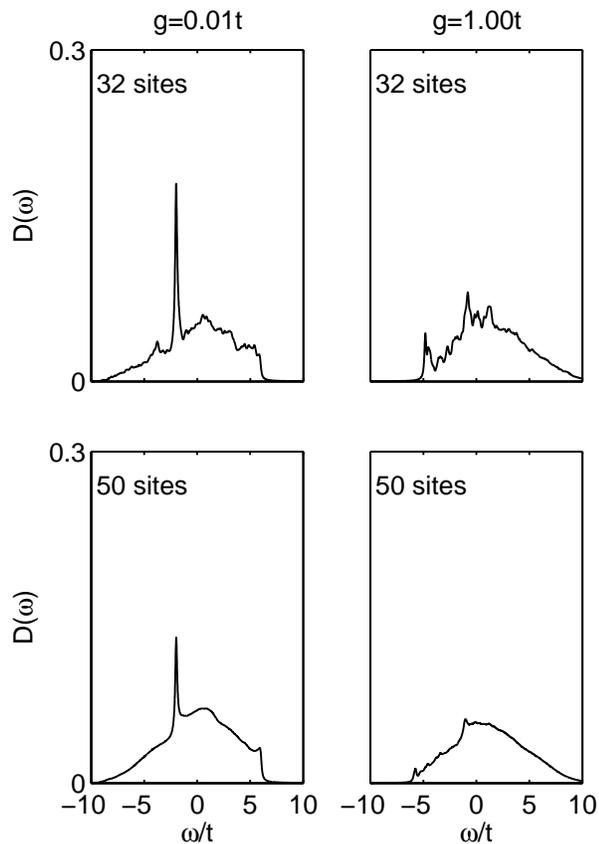}\end{center}

\caption{Integrated spectral density of a 32 sites cluster calculated for
the effective $t-g$ Hamiltonian for two different values of $g$.
The peaks are broadened by choosing $\eta=0.1t.$\label{cap:Spectral-density_Heff} }
\end{figure}

The integrated spectral density $D(\omega)$ of the effective Hamiltonian
(\ref{eq:Heff}) shows a strong dependence on the ratio $t/g$. In
the {}``physical'' regime, corresponding to the previously considered
parameters with $g\sim t^{3}/V^{2}=0.01t$, the integrated spectral
density $D(\omega)$ (Fig. \ref{cap:Spectral-density_Heff}) shows
qualitatively the same features as the full Hamiltonian. If $g$ is
assumed to be equal to $t$, the broad continuum at the bottom of
the spectral density vanishes and a sharp $\delta$-peak evolves instead.
Note the similarities to $D(\omega)$ for the artificially confined
situation. The spectral weight of the peak can be looked at as that
of a Landau quasiparticle peak. This suggests the interpretation that
the ring exchange term leads to charge order which is destroyed by
the separation of the two fcp's. In the regime with $g\ll t$ ($g=t^{3}/V^{2}$)
the diameter of the two bounded fcp's is larger than the considered
system size and thus the excitations seem to be deconfined. An artificial
increased $g$ leads to much stronger confinement and the diameter
of the bounded pair is small compared to the system size - leading
to a finite weight of the quasiparticle peak. The huge spatial extend
of the quasiparticle in the {}``physical'' regime is expected to
lead to very interesting effects, but this will be the subject of
a separate investigation.

\subsection{Optical Conductivity}

The regular part of the optical conductivity $\sigma_{\mbox{\tiny reg}}$
is defined by \begin{eqnarray}
\sigma_{\mbox{\tiny reg}}(\omega) & = & \lim_{\eta\rightarrow0^{+}}-\frac{1}{\omega\ \pi}\\
 & \times & \mbox{Im}\langle\psi_{0}^{N}|j_{\mathbf{x}}\ \frac{1}{\omega+i\eta+E_{0}-H}\  j_{\mathbf{x}}|\psi_{0}^{N}\rangle.\nonumber \end{eqnarray}
Here $|\psi_{0}^{N}\rangle$ is the ground state of the system with
$N$ particles and as before we use a value of $\eta=0.1$ for broadening.
The current operator is\begin{eqnarray}
j_{\mathbf{x}} & = & i[H,X]\nonumber \\
 & = & it\sum_{j}\left(c_{\mathbf{r}_{j}}^{\dag}c_{\mathbf{r}_{j}+\mathbf{x}}^{\vphantom{\dag}}-c_{\mathbf{r}_{j}+\mathbf{x}}^{\dag}c_{\mathbf{r}_{j}}^{\vphantom{\dag}}\right).\label{eq:current}\end{eqnarray}
Calculations have been performed for the current density along the
$x$-axis, i.e. $\mathbf{x}=\mathbf{e}_{x}$ in \ref{eq:current}.%
\begin{figure}
\begin{center}\includegraphics[%
  width=80mm]{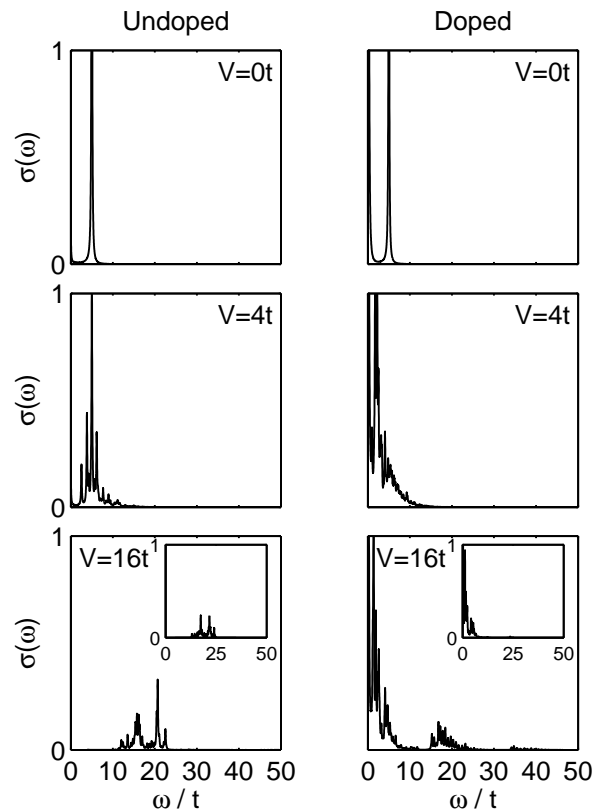}\end{center}

\caption{Regular part of optical conductivity $\sigma_{\mbox{\tiny reg}}(\omega)$
of the $\sqrt{18}\times\sqrt{18}$ checkerboard lattice for increasing
nearest-neighbor repulsion $V$. Left panels are showing the half
filled case and the right half-filled plus one. The insets show the
data for the same values within the minimal Hilbert space. The peaks
are broadened by choosing $\eta=0.1t.$\label{cap:Optical-conductivity_3x3}}
\end{figure}
 The regular part of the optical conductivity $\sigma_{\mbox{\tiny reg}}(\omega)$
is shown in Figure \ref{cap:Optical-conductivity_3x3} for increasing
values of the nearest-neighbor repulsion $V$. For $V=0,$ one observes
a peak-like structure. In the thermodynamic limit it should be positioned
at $\omega=0$ and the shift to $\omega\neq0$ is a finite size effect.
The same holds true for V=4t, which is expected to be still in the
metallic phase.\cite{runge2004} With increasing $V,$ the structures
become broader. In the half-filled (undoped) case the complete weight
is moved to larger $\omega$ and the Drude weight goes to zero. One
expects a transition to an insulating state. For large $V$ the weight
is distributed around $\omega=V$. This corresponds to the energy
that is needed to generate a fcp-fch pair which carry an electrical
current. 

In the doped case one finds a different behavior. Finite weight is
found at $\omega=0$ for arbitrarily large values of $V$ and the
charge is carried by two fcp's with charge $e/2$ each. A part of
the weight is shifted to larger $\omega$ where additional fcp - fch
pairs are generated as charge carriers. For $V=16t$ we compare the
results based on the calculation in the full Hilbert space with those
for the minimal Hilbert space. The features at small $\omega$ are
reproduced very well, but in the minimal Hilbert space vacuum fluctuations
are absent and thus there is no current contributions from fcp-fch
pairs.

\begin{figure}
\begin{center}\includegraphics[%
  width=80mm]{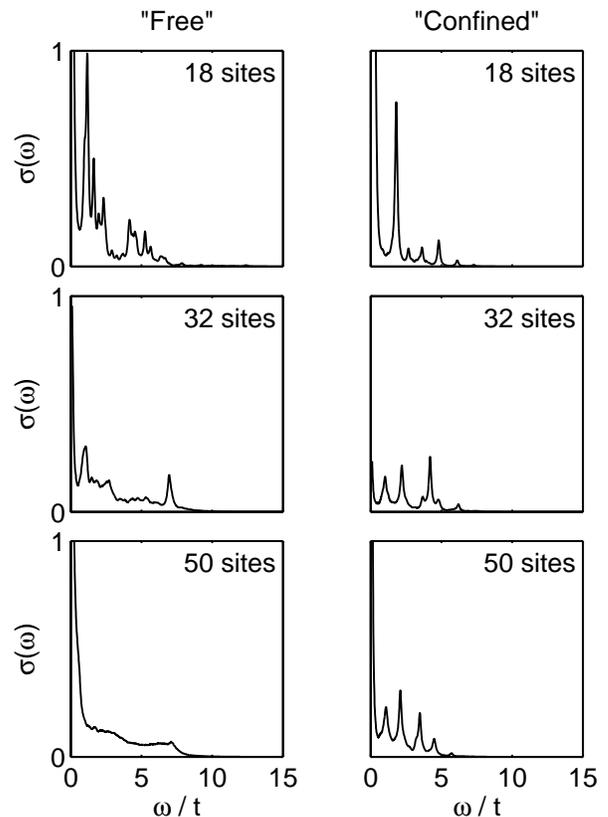}\end{center}

\caption{Regular part of optical conductivity $\sigma_{\mbox{\tiny reg}}(\omega)$
for the doped system with $V=25t$ calculated for different lattice
sizes. The left panel shows the data for {}``Free'' systems which
include states with separated fractional charges. The right panel
shows the data for systems where the two defect are {}``confined''
and can not separate. The peaks are broadened with $\eta=0.1t.$\label{cap:Regular-part_V25t} }
\end{figure}

Now we want to investigate the influence of fractional charges on
the optical conductivity (see Fig. \ref{cap:Regular-part_V25t}).
The main features are independent of lattice size. For the {}``free''
system one finds a bandwidth of nearly $13t$ which is the same as
for the spectral functions. There are sharp peaks superposed onto
a broad structure. The {}``confined'' system has a reduced bandwidth
of nearly $8t$ and the broad structure is less pronounced. For the
32 sites cluster the contribution near $\omega=0$ is smaller than
for the other two clusters. This is due to the particular shape of
the clusters which lead to a lower degeneracy of the ground state
than in the 32 sites case. 

\begin{figure}
\begin{center}\includegraphics[%
  width=80mm]{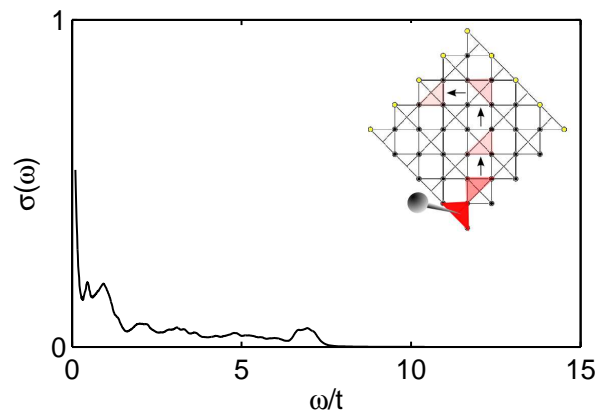}\end{center}

\caption{Regular part of optical conductivity $\sigma_{\mbox{\tiny reg}}(\omega)$
for the doped system with $V=25t$. Only one of the fractional charged
excitations is mobile while the other is fixed. The peaks are broadened
by choosing $\eta=0.1t.$ Arrows mark motions of particles.\label{cap:Regular-part_V25t_one_free} }
\end{figure}

In Figure \ref{cap:Regular-part_V25t_one_free} %
the optical conductivity due to \emph{one} moving fcp is shown. Here
we keep one of the two fcp's fixed and let only the other one propagate.
One observes a Drude peak and broad structure. Except for the Drude
peak no other peaks are observed. The bandwidth is reduced to about
half of the bandwidth of two fcp's.

\section{Summary and outlook\label{sec:Summary-and-outlook}}

We have studied numerically the dynamical properties of spinless fermions
on checkerboard lattices and compared the results for different lattice
sizes and models. For the full Hamiltonian a broad low-energy continuum
is found and no Fermi liquid peak is present, indicating that an added
particle decays into two fractionally charged excitations that separate
over the whole finite lattice. 

Considering the ring exchange $g$ in the effective Hamiltonian as
an independent parameter (not fixed to $t^{3}/V^{2}$) allowed us
to explore the regime where the fcp separation is small compared to
the system size. The spectral function for large $g$ does not show
a broad structure and develops instead a sharp peak. The existence
of a quasiparticle peak shows that the added particle creates two
bound fractionally charged particles with a small diameter. These
findings suggest that for parameters $V/t\approx10$ quasiparticles
with spatial extend over more than hundred lattice sites are formed. 

Finally the questions about the dynamics of fractional charges on
different types of lattices remains. Considerable differences are
expected for bipartite and non-bipartite lattices. Also the dimensionality
of the lattice is crucial. For the 3D pyrochlore lattice one speculates
about the existence of liquid phases where the fractional charges
would be deconfined.\cite{hermele2004,moessner2003a,fulde2002} The
informations we got from investigations of finite checkerboard lattices
suggest that the possibly deconfined fcp's in the pyrochlore lattice
would lead to broad features in spectral functions even for infinite
systems.

\subsection*{Acknowledgments }

We thank Professor Claire Lhuillier and Professor Joseph Betouras for stimulating
discussions and helpful remarks. One author (F. P.) has benefited
from the hospitality of IDRIS, Orsay during a stay made possible by
a HPC Europe grant (RII3-CT-2003-506079) . 

\bibliographystyle{apsrev}
\bibliography{./mainbib}

\end{document}